Communication

# Spin-related electronic pathway through single molecule on Au(111)

Mingjun Zhong[a], Qimeng Wu[a], Liang Ma[b,] *, Jie Li[a], Yifan Wang[a], Yansong Wang[a], Xin Li[a], Yajie Zhang[a], Jingtao Lü[c,] *, Yongfeng Wang[a,d,e] *

[a]Center for Carbon-based Electronics and Key Laboratory for the Physics and Chemistry of Nanodevices, Department of Electronics, Peking University, Beijing 100871, China
[b]Key Laboratory for Microstructural Material Physics of Hebei Province, School of Science, Yanshan University, Qinhuangdao 066004, China
[c]School of Physics, Institute for Quantum Science and Engineering, and Wuhan National High Magnetic Field Center, Huazhong University of Science and Technology, Wuhan 430074, China
[d]Beijing Academy of Quantum Information Sciences, Beijing 100193, China
[e]Institute of Spin Science and Technology, South China University of Technology, Guangzhou 511442, China

ABSTRACT

Spin properties of organic molecules have attracted great interest for their potential applications in spintronic devices and quantum computing. Fe-tetraphenyl porphyrin (FeTPP) is of particular interest for its robust magnetic properties on metallic substrates. FeTPP is prepared in vacuum via on-surface synthesis. Molecular structure and spin-related transport properties are characterized by low-temperature scanning tunneling microscope and spectroscopy at 0.5 K. Density functional theory calculations are performed to understand molecular adsorption and spin distribution on Au(111). The molecular structure of FeTPP is distorted upon adsorption on the substrate. Spin excitations of FeTPP are observed on the Fe atom and high pyrrole groups in differential conductance spectra. The calculated spin density distribution indicates that the electron spin of FeTPP is mainly distributed on the Fe atom. The atomic transmission calculation indicates that electrons transport to substrate is mediated through Fe atom, when the tip is above the high pyrrole group.

The spin-related properties of organic molecules are of great significance not only in fundamental science but also in constructing novel organic molecular spintronic devices like spin transistors and spin valves with low energy cost and long spin relaxation time [1][6]. According to previous reports, by evaporating magnetic organic molecules or atoms onto substrate, their intrinsic magnetic properties can be studied by scanning tunneling microscopy (STM) [7][13]. Generally, organic molecules may have structure distortions upon adsorption on surface due to their mutual interaction [19], and this kind of distortion may influence the magnetic properties of molecules [20],[21]. However, the specific mechanism of molecular distortions affecting the magnetic properties of molecules still needs to be clarified.

Fe-tetraphenyl porphyrin (FeTPP) is one of the representative organic magnetic molecules for its robust magnetic properties on metallic substrates [22]. The porphyrin ring of FeTPP on Au(111) has a saddle conformation with two opposite pyrrole groups bending up (high pyrrole groups ) and the other two bending down (low pyrrole groups). Therefore, FeTPP is an ideal model molecule to study spin-related electronic pathway through single non-planar molecule on substrates by STM. The transport properties of FeTPP/Au(111) are calculated by the atomic transmission using nonequilibrium Green's function method combined with density function theory to compare with the experimental results.

All experiments are performed in an ultrahigh vacuum low temperature STM, and the d$I$/d$V$ spectra are measured at 0.5 K with the lock-in amplifier technology. The modulation frequency is set at 967 Hz and the modulation voltage is 0.2 mV. The Au(111) surface is cleaned by circles of Ar$^+$ sputtering and annealing at around 800 K. Due to the instability of FeTPP under atmospheric condition [23], the molecule is in-situ synthesized by evaporating 2HTPP molecules and Fe atoms onto Au(111) and annealing the substrate. The spin density of FeTPP adsorbed on four-layer Au(111) slab is calculated by density functional theory (DFT) as implemented in the Vienna *ab initio* simulation package (VASP) [24]. The generalized gradient approximation developed by Perdew-Burke-Ernzerhof (PBE-GGA) is used to describe the exchange correlation potential [25]. A cutoff energy of 400 eV for the plane-wave basis set is chosen and Γ-point approximation is used. The bottom two layers of the substrate are fixed during the structural optimization. The structure is relaxed until the force acting on each atom is less than 0.02 eV/Å. To avoid interactions between neighboring cells, the slabs are separated by a 10 Å vacuum in the z direction. The atomic transmission through single FeTPP molecule is calculated using the package TranSIESTA [26],[27].

---

* Corresponding author.
 *E-mail address*: maliang-phy@ysu.edu.cn
* Corresponding author.
 *E-mail address*: jtlu@hust.edu.cn
* Corresponding author.
 *E-mail address*: yongfengwang@pku.edu.cn

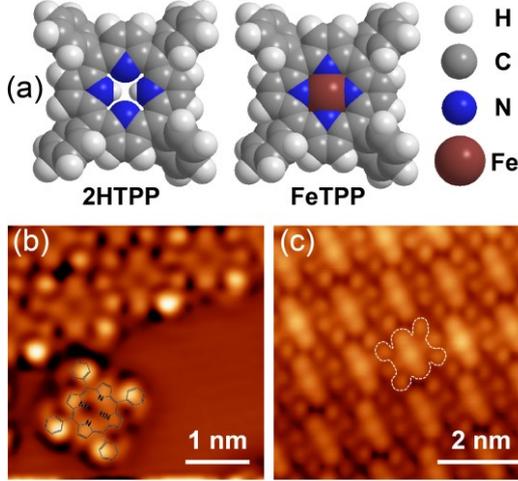

**Fig. 1.** (a) Chemical structures of 2HTPP and FeTPP. (b) STM image of 2HTPP molecules on Au(111) ($V = 20$ mV; $I = 20$ pA). A molecular model of 2HTPP is overlaid on its STM image. (c) STM image of close-packed FeTPP on Au(111) ($V = 20$ mV; $I = 20$ pA). The white dashed line shows the contour of a single FeTPP molecule.

The chemical structures of 2HTPP and FeTPP are shown in Fig. 1a. After being deposited onto Au(111) surface, the center of 2HTPP appears dark in the STM image, as shown in Fig. 1b. Further depositing Fe atoms and annealing the substrate leads to the formation of FeTPP molecules. STM image of FeTPP molecules is shown in Fig. 1c. The white dashed line represents the contour of a single FeTPP molecule with a saddle conformation, just like the previous reports [28]-[30]. Compared with 2HTPP, a protrusion is observed at the center of FeTPP.

In order to study the spin characteristics of FeTPP, $dI/dV$ spectra are measured on molecular different positions. The step-like features are observed in the spectra measured on Fe atom and bright pyrrole groups, as shown in Fig. 2a and Fig. 2b, respectively. However, the spectra measured on the dark pyrrole group are featureless. These results are consistent with the previous reports, and the step-like signals originate from spin excitations [28].

DFT calculations are performed to understand the spin-related transport of FeTPP on Au(111). The magnetic moment of FeTPP is 2.1 $\mu_B$, indicating that FeTPP is a spin S = 1 system. The spin excitation process at the Fe position can be simply described by the following spin Hamiltonian $H_{eff} = DS_z^2 + E(S_x^2 - S_y^2)$. Here, D and E represent the axis anisotropy and transverse anisotropy, respectively, and $S_x$, $S_y$, $S_z$ refer to the spin operators at different directions. D and E can be extracted from the spectrum by the methods proposed by Markus [29], and are found to be 7.9 meV and 0.92 meV for the spectrum measured on Fe atom (Fig. 2a). The nonzero D and E will generate the zero-field splitting (ZFS), resulting in double steps in $dI/dV$ spectra. The diagram of energy level splitting and excitation process is shown in Fig. 2c. According to the diagram, the first step in $dI/dV$ spectrum refers to the spin excitation from $|\psi_0\rangle$ to $|\psi_1\rangle$, and the second step refers to the spin excitation from $|\psi_0\rangle$ to $|\psi_2\rangle$). The differential conductance measured on Fe atom is larger at negative bias. The continuous spectra measured along the two opposite high pyrrole groups, which are marked by the dark line in the inset of Fig. 2a, indicate that the asymmetric feature is mainly located at the Fe atom (Fig.2d). This asymmetry in spectra is attributed to potential scattering during tunneling process [29],[30].

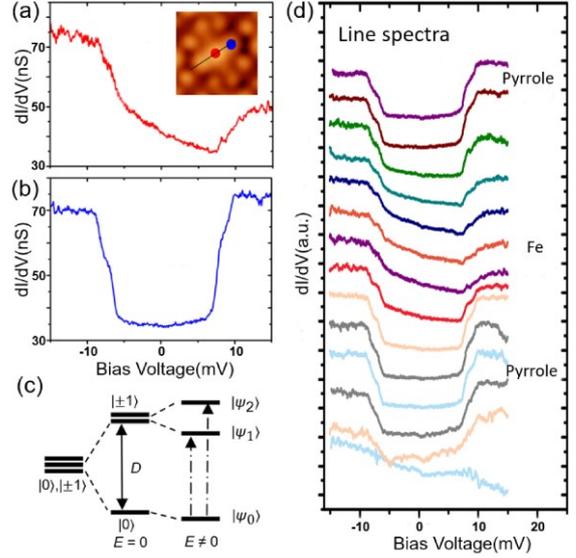

**Fig. 2.** (a-b) D$I/dV$ spectra measured on Fe atom and high pyrrole group of FeTPP, respectively. The specific measured positions on Fe atom and high pyrrole are marked by red and blue dots in the inset of (a). (c) Diagram of ZFS of FeTPP. Due to the nonzero D and E, the initial degenerate states split into $|\psi_0\rangle$, $|\psi_1\rangle$, and $|\psi_2\rangle$ states. (d) Line spectra measured along the dark line in the inset of (a). Each spectrum is vertically offset for clarity.

The next step is to understand the mechanism of spin excitation at the high pyrrole group in spectra. The spin distribution of FeTPP is calculated and the electron spin is mainly localized at Fe atom (Figs. 3a and 3b), where molecular four phenyl groups are omitted from the plots for clarity. This result suggests that the spin signal at pyrrole should be related to the Fe atom, which is confirmed by further geometric and transport calculations. Fig. 3b presents the side view of the optimized structure of FeTPP on Au(111). It is clear that two pyrrole groups are much higher than the others. The distance between nitrogen atom in the high pyrrole group and substrate is 0.34 Å higher than that between nitrogen atom in the low pyrrole group and substrate. Two high pyrrole groups appear brighter in STM images. It is reasonable that electrons move from the bright pyrrole group to Fe and then to substrate instead of tunneling to substrate directly, when the STM tip locates above bright pyrrole.

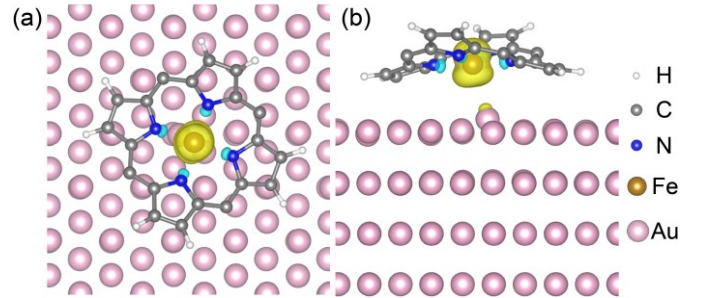

**Fig. 3.** (a-b) Top and side views of spin-density distribution of FeTPP on Au(111). Molecular four phenyl groups are omitted from the plots for clarity.

The atomic transmission through FeTPP is calculated to quantitatively understand electronic pathway. We mainly focus on the transmission of Fe atom with different tip positions. The schematic of the FeTPP-based tunneling junction is shown in Fig. 4a. The simulated tip positions are shown in Fig. 4b, represented by the terminated Au atom of the tip. When calculating the atomic transmission with the tip above the Fe atom, the simulated tip slightly deviates from the right above position of Fe atom to keep the structure of the tip. The transmissions through Fe and other atoms are shown in Fig. 4c, where the sky-blue, pink and green columns refer to the atomic transmission fractions with the tip on Fe atom, high and low pyrrole groups, respectively. When the tip is above the Fe atom, most electrons tunnel through Fe to substrate directly (90%). When the tip is placed above the high pyrrole group, the simulation shows that electrons prefer to transport through Fe atom as well (96%). However, when the tip is above the low pyrrole groups, only a small part of electrons (28%) transport through Fe atom.

When the tip is on high and low pyrrole groups, electrons transport through Fe atom, indicating the existence of lateral transport process in molecular skeleton from pyrrole groups to Fe atom. The different transmission fractions through Fe atom demonstrate that the structure distortion of FeTPP influences the contribution of different electron transport path. When the tip is on the high pyrrole, electrons tunnel to pyrrole firstly, and then transfer to Fe atom. When tunneling from Fe to Au(111), the spin state of Fe atom is excited, which leads to the inelastic step-like features in spectra. Moreover, compared with the spectra measured on Fe atom, the spectra measured on the high pyrrole group are more symmetric because the potential scattering interaction does not exist between high pyrrole and tip. When the tip is on low pyrrole group, electrons tunnel through Fe atom as well but with a small fraction, which might account for the absence of spin excitation in spectra.

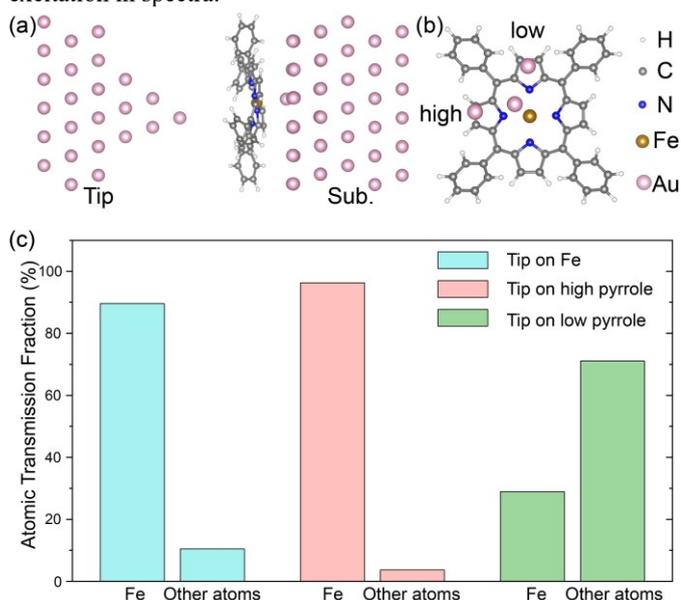

**Fig. 4.** (a) Chemical model of a tip-FeTPP-substrate tunneling junction. (b) Three different tip positions used to calculate the atomic transmission. Only the terminated Au atom of the tip is used to represent the tip position. High pyrroles are on the left side and right side of the structure. (c) Atomic-transmission fractions with different tip positions. The sky-blue, pink and green columns refer to the atomic transmission fractions with the tip on Fe atom, high pyrrole and low pyrrole, respectively.

In summary, spin-related electronic pathways through single FeTPP molecule on Au(111) are determined by STM, DFT and atomic transmission calculations. FeTPP adsorbs on the substrate in a saddle conformation with two pyrrole groups bending towards vacuum and the others bending to the substrate. Spin excitations of FeTPP are found on both Fe atom and high pyrrole groups in differential conductance spectra. DFT calculations indicate that molecular electron spin is localized at the Fe atom and total spin $S = 1$, which is responsible for the spin excitation in spectra. The atomic transmission calculation indicates that electrons transport to substrate is mediated through Fe atom. When the tip is on the high pyrrole group, electrons prefer to transport through Fe atom to substrate instead of tunneling from high pyrrole to substrate directly.

## Acknowledgments

This work is supported by the Ministry of Science and Technology (Nos. 2017YFA0205003, 2018YFA0306003) and National Natural Science Foundation of China (Nos. 21991132, 21972002, 21902003, 21673118, 21972067).